# The effect of shock-wave profile on dynamic brittle failure

J.P. Escobedo[1], E.N. Brown[2], C.P. Trujillo[1], E. K. Cerreta[1] and G.T. Gray III[1]

[1]MST-8, MS G-755, Los Alamos National Laboratory, Los Alamos, NM 87545, U.S.A.
[2]P-23, MS H-803, Los Alamos National Laboratory, Los Alamos, NM 87545, U.S.A.

**Abstract.** The influence of shock-wave-loading profile on the failure processes in a brittle material has been investigated. Tungsten heavy alloy (WHA) specimens have been subjected to two shock-wave loading profiles with a similar peak stress of 15.4 GPa but different pulse durations. Contrary to the strong dependence of strength on wave profile observed in ductile metals, for WHA, specimens subjected to different loading profiles exhibited similar spall strength and damage evolution morphology. Post-mortem examination of recovered samples revealed that dynamic failure for both loading profiles is dominated by brittle cleavage fracture, with additional energy dissipation through crack branching in the more brittle tungsten particles. Overall, in this brittle material all relevant damage kinetics and the spall strength are shown to be dominated by the shock peak stress, independent of pulse duration.

**I Introduction**

It has been well established that dynamic fracture or spall is a complex process strongly influenced by both the microstructure and the loading profile — the shape of the shock wave as a function of time—imparted to the specimen[1-9]. For ductile metals it has been observed that the spall response is highly dependent on the pulse shape in addition to the peak stress. For instance, the shock response in copper has been observed to be very sensitive to the loading profile[9, 12]. While a square-wave with a peak stress of 8 GPa resulted in localized damage with sufficient ductile void coalescence that caused complete separation of a single spall scab, loading with triangular wave at the same peak stress revealed a consistent ringing in the particle velocity indicating formation of a free surface or a significant damage layer in the material but no voids were formed. Furthermore, the spall strength for the square and triangular wave loading were reported to be 1.28 and 2.04 GPa, respectively[12]. Similarly, Gray *et al*[10, 13] showed for 316L stainless steel that while loading with a square-wave resulted in incipient spallation in 316L SS at a peak stress of 6.6 GPa, the material loaded with a triangular wave exhibited no "pull-back" nor any damage evolution until a peak stress of at least 14.5 GPa. Mechanistically, the dependence on loading profile is coupled to the kinetics of ductile plastic deformation leading to void nucleation, growth, and subsequent linkage to form a spall plane[1, 5]. As such, sufficient energy—supplied by the shock pulse—is required to be applied over sufficiently long times to accommodate the kinetics of the deformation and damage evolution mechanisms.

In this regard, the effect of shock wave profile on spall strength is analogous to the effect of frequency and wave profile (typically sinusoidal, triangular, or square) on fatigue crack growth rate [11, 14]. Shock loading is effectively a single cycle of compression followed by tension during which damage is accumulated. According to the classic Paris law, the fatigue crack growth rate, as defined by the length of crack extension per loading cycle, is dependent on the stress intensity factor range, as defined from the maximum and minimum loading states[15]. This simple relationship ignores the details of how the cyclic stress is applied, as such, the applicability of this assumption has been shown to be material dependent. For example, glassy brittle polymers, such as polycarbonate[16], have been shown to exhibit negligible frequency dependence, while ductile polymers such as polyvinylchloride[16] or polymethylmethacrylate[17] exhibit a strong dependence on frequency with a notable decrease in fatigue crack growth rate for a given stress intensity factor range with increasing frequency. The rationale is that for a brittle material the time for incremental crack growth to occur is effectively instantaneous compared to the cycle time, making frequency and wave profile inconsequential. A ductile material on the other hand requires time to accommodate the kinetics of the deformation and damage mechanisms, as void growth is a volume-additive process and therefore time dependent, thus requiring many more cycles as the frequency increases to obtain comparable time at stress. This observation of the interplay between the kinetics of damage mechanisms and the frequency and wave profiles effects in fatigue crack growth inspired us to consider the effect of shock wave profile in classes of materials known to differ in kinetic response, such as the difference between a ductile and a brittle metal [9-12]. Having previously considered ductile materials with damage and deformation kinetics that are slow relative to the shock wave, here we consider a brittle material with damage and deformation kinetics that are fast relative to the shock wave. Specifically, in the present study we investigate the failure processes in a brittle tungsten heavy alloy.

Tungsten heavy alloy (WHA) is a composite material of tungsten particles (body-centered-cubic) in an austenitic matrix (face-centered-cubic) comprised of tungsten, nickel, and iron. While the composite is known to fail in a brittle manner, it displays improved ductility under compression and tension in comparison to that of pure polycrystalline tungsten[18]. The shock response has been reported in the literature for pure polycrystalline tungsten [18-26] and tungsten heavy alloys[27-36] under different loading environments. For instance, in plate impact experiments using monolithic impactors, Zurek and Gray [18] reported spall strengths of WHA to be 3.4-3.8 GPa. Dandekar and Weisgerber[30] reported the spall strength of WHA to be 1.7–2.0 GPa, whereas Bless and Chau[32, 33] reported the spall strength for WHA to be 2.6 GPa. Vogler and Clayton[34] employed line-VISAR to spatially resolve statistics on the spall strength of WHA. By fitting their measurements to both normal and Weibull distributions, they reported the average spall strength within a given sample varied from 1.7 to 2.1 GPa with standard deviations from 0.3 to 0.6 GPa. On the other hand, Chang and Choi[37] employed RDX to explosively load WHA, imparting a Taylor-wave-loading profile. They reported spall strength values of WHA as high as 5.6 GPa, significantly higher than those obtained in plate impact experiments using monolithic flyers. However, Baoping et al.[38] also employed explosive loading and reported spall strength values of 0.5 to 3.0 GPa, which are within the reported values in the literature for plate impact experiments.

In addition to the ambiguity on the effect of loading profile on the shock response of WHA, most of the cited work has been focused specifically on assessing the shock response via interferometry measurements at the free surface, i.e. spall strength. However, this value represents many intertwined parameters and fracture mechanisms that precludes any significant conclusions about the kinetics of loading on brittle failure mechanisms. To address this problem, the main objective of the present study is to elucidate the effect of loading profile on the fundamental mechanisms of brittle fracture (crack nucleation and propagation) in WHA specimens. To this end, spall experiments are performed on WHA samples with two significantly distinct shock loading profiles, i.e. pulse durations and accompanying unloading rates. Detailed fractographic analyses of the damage in the spalled WHA samples as a function of shock-wave profile of comparable peak stress is presented. For both profiles, it is observed that the failure in WHA is by brittle trans-particle crack growth with additional energy dissipation through crack branching in the more brittle tungsten particles. Additionally, we also observe that for the current peak shock stress (15.4 GPa), the wave profile does not influence the spall strength significantly. This is believed to be directly linked with the relative insensitivity of WHA to time dependent processes.

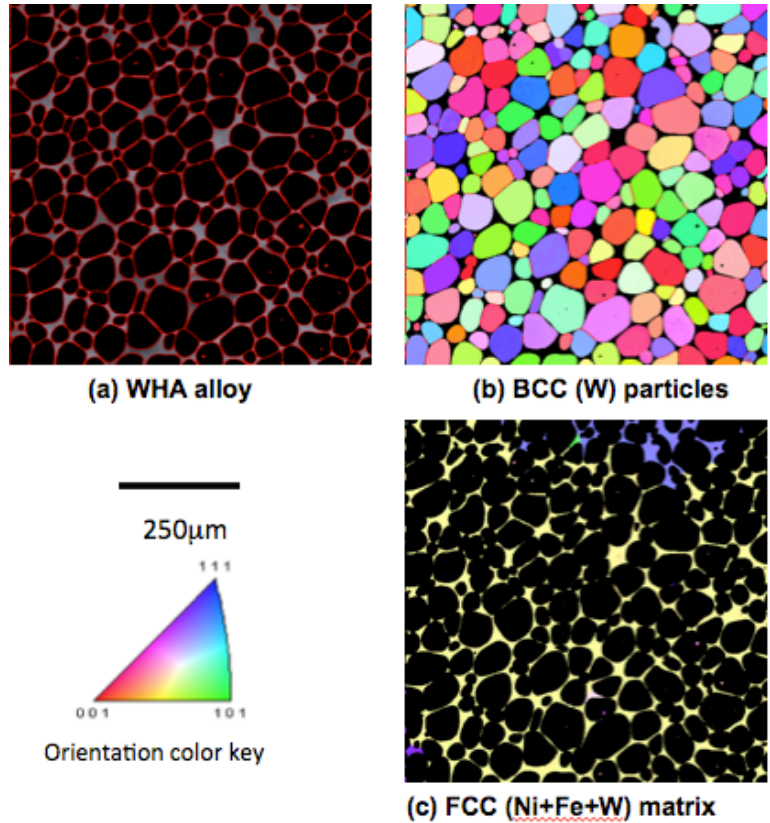

**Fig. 1** EBSD results of initial characterization. (a) Phase map of the WHA alloy. Dark phase correspond to the W particles, the light-colored phase is the matrix, and the red lines demarcate particle –matrix interfaces. (b) Orientation map showing only the W-particles. No texture is observed. (c) Orientation map showing only the large crystalline domains of the FCC matrix. For the last two maps, the crystalline orientation color is according to the color key.

**II Experimental methods**
**A. Initial Material Characterization**

All target materials were prepared from a tungsten heavy alloy (WHA) plate with a composition of ~92.5% W and the remainder of Ni and Fe. The alloy possesses a composite microstructure of tungsten particles in an austenitic matrix comprised of tungsten, nickel and iron. Representative results of the initial characterization, performed by means of electron backscatter diffraction (EBSD) measurements, are given in Figs 1.(a)-(c). A phase map of the metallic composite is given in Fig 1.(a). The dark regions correspond to the W particles, identified as a phase with a body centered cubic (BCC) structure. The red lines demarcate the boundary between the W particles and the matrix. The matrix is shown as the light-colored phase, identified as a phase with a face centered cubic (FCC) structure. An orientation map showing only the W particles is given in Fig 1.(b). The color of each particle is assigned based on its specific crystallographic orientation, as given by the accompanying color key. No preferred orientation, i.e. texture, is observed in the W particles. The orientation map showing the FCC matrix is shown in Fig 1.(c). Large FCC crystalline domains are observed to dominate the microstructure in the matrix. To obtain a statistical representation of the distribution of the W particles, additional maps like those shown in Figs 1.(a)-(c) were generated at different areas of the as-received WHA plate. These measurements yielded values for the area fraction (i.e. volume fraction) of W particles of $0.859 \pm 0.003$ and a particle size of $\sim 50 \pm 19$ μm. In addition, the particle-matrix boundary density was measured to be $86.2 \pm 1.9$ mm/mm$^2$, while the particle-particle boundary density was $10.3 \pm 0.3$ mm/mm$^2$.

**Table I:** Parameters for the plate impact experiments.

| Exp. ID | Impactor | | | Peak free surface velocity (mm/μs) | Pulse duration (μs) | Compressive stress (GPa) |
|---|---|---|---|---|---|---|
| | Material | Thickness (mm) | Velocity (mm/μs) | | | |
| L (Long Pulse) | W | 2.0 | 0.383 | 0.375 | 0.5 | 15.37 |
| S (Short Pulse) | W/ldm | 0.5/5 | 0.391 | 0.377 | 0.05 | 15.46 |

Notes: (a) ldm = low density material, microballoons

**B. Plate Impact Experiments**

Plate impact experiments were conducted using the 80 mm bore gas launcher previously presented by Gray[39]. Two identical WHA targets, 20mm in diameter and 4mm-thick, were prepared with press fit momentum trapping rings to mitigate perturbations from edge release waves. The experimental parameters are listed in Table I and the schematics for the two experiments are shown in Fig. 2. A loading profile with a long-pulse duration was achieved by using a 2 mm-thick W monolithic impactor. From here on, this long-pulse loading profile is referred to as profile L. The relative thickness of the target and impactor was chosen to locate the spall plane at the midline of the target (Fig 2.(a)). Furthermore, this geometry causes a relatively slow interaction of release waves that generates a wide tensile pulse within the WHA. Alternatively, a profile with a significantly shorter pulse duration was achieved with a layered impactor consisting of a 0.5mm-thick W backed with a low-density microballoon composite ($\rho = \sim 400$ kg/m$^3$, almost 50 times lower density than WHA). From here on, this second short-pulse profile is referred to as profile S. For this geometry, the release waves interact relatively faster than the previous case which results in a narrower tensile pulse (Fig 2.(b)). Impact velocity was measured to an accuracy of 1% using a sequential pressure transducer method and sample tilt was fixed to ~1 mrad by means of an adjustable specimen mount. Both experiments were executed to achieve a peak compressive stress of ~15.4 GPa, which is significantly higher than the reported range for the spall strength of WHA. The free surface velocity (FSV) profiles were measured using Photon Doppler Velocimetry (PDV)[40, 41] single-point probes. Following impact, all samples were soft recovered by decelerating them into low-density foam.

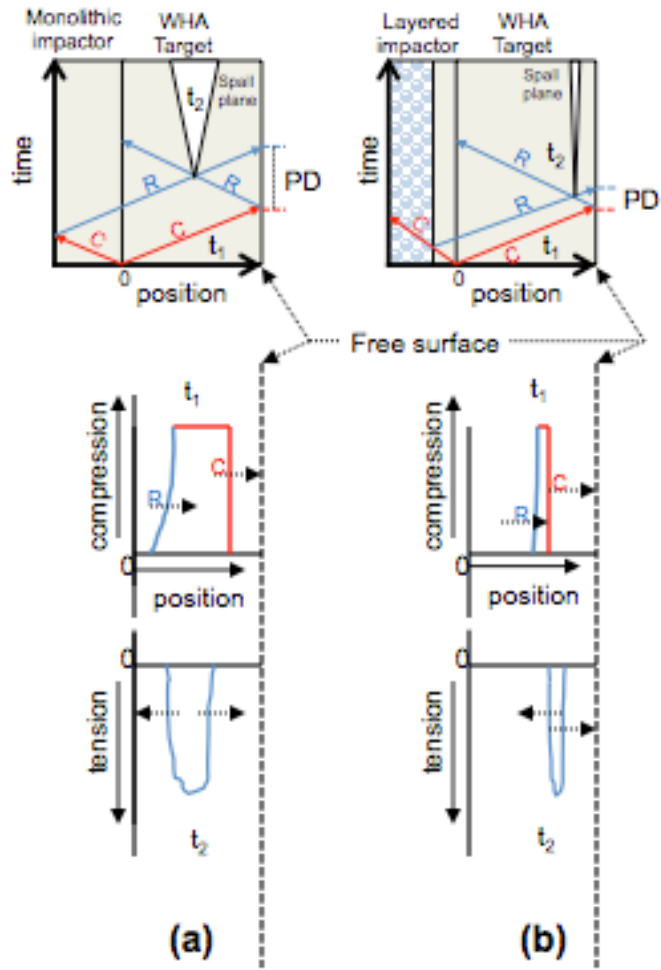

**Fig. 2** Schematics showing the x-t diagrams of the trajectory and interaction of compressive (C, colored as red) and release (R, colored as blue) waves, as well as the development of compressive and tensile pulses for: (a) the monolithic WHA impactor imposing a loading profile with long pulse duration in compression and wide tensile pulse; and the (b) the layered impactor imposing a loading profile with shorter pulse duration in compression and a narrow tensile pulse.

### C. Post-impact metallurgical characterization

Examination of the damage fields in the soft recovered spalled samples included optical, scanning electron microscopy (SEM) as well as electron backscatter diffraction (EBSD). In preparation for the metallographic analyses, each recovered specimen was diametrically sectioned. The sections were then mounted in an epoxy resin and prepared following standard metallographic techniques up to a 0.05 µm colloidal silica final polish, performed on a vibratory polisher. Chemical etching was performed at intermediate polishing steps using a solution of 100ml. of $H_2O$, 15g of $K_3Fe(CN)_6$ and 2g of NaOH. A similar procedure was followed for the as-received WHA plate. Optical microscopy was performed on a Zeiss microscope. SEM and EBSD microscopy were performed on a Phillips XL30 FEG equipped with a Hikari high speed detector. The instrument was operated using a voltage of 20kV, with step sizes of 0.5-1.5µm, depending on the feature of interest. Data was acquired and analyzed using orientation imaging microscopy (OIM) software by TexSEM Laboratories (TSL) of EDAX. The OIM software aided to identify crystallographic orientations, as well as with the calculation of properties obtained from EBSD measurements such as Elastic Stiffness maps.

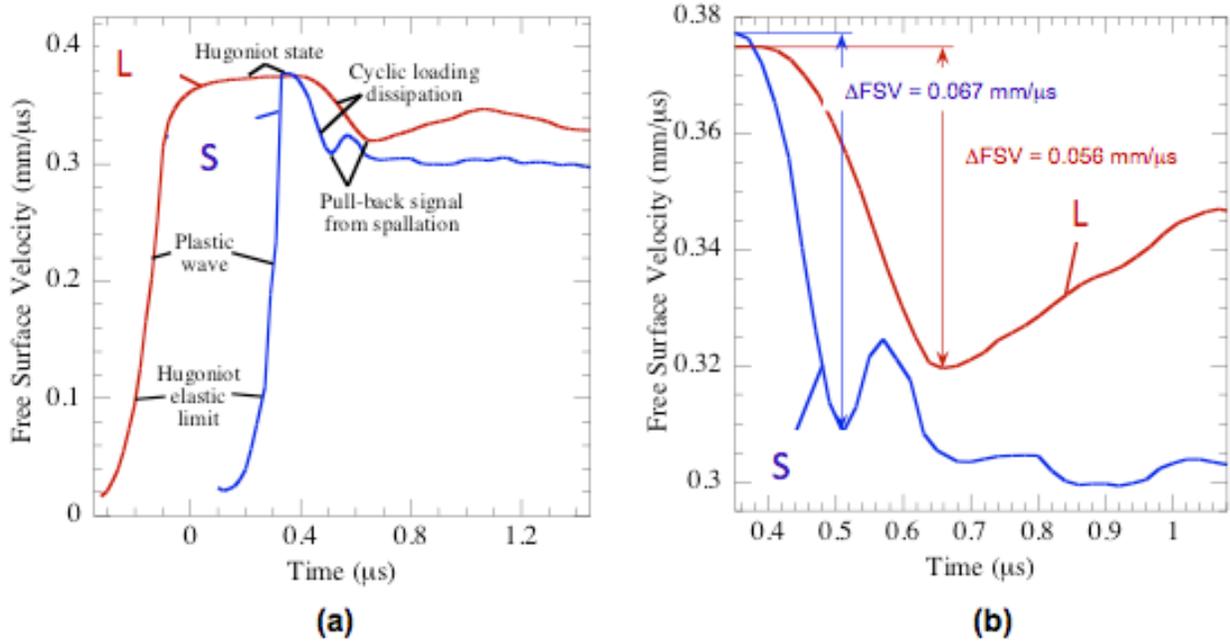

**Fig. 3** (a) Traces of the free surface velocity showing shock loading with long (L) and short (S) pulse durations. The curves are plotted such that the drop from the peak state starts at similar times for both experiments. (b) Region of the free surface velocity highlighting the pull-back signal, indicative of spall plane formation.

## III Results and Discussion
### A. Free surface velocity profiles

The free surface velocity (FSV) profiles for the experiments reported herein are shown in Fig 3.(a). These measurements were performed at the sample back free surface. Key parameters and calculated values are listed in Table I. The peak free surface velocities are 0.375 and 0.377mm/μs, corresponding to peak compressive stresses of ~15.4GPa (calculated using the Hugoniot parameters for tungsten: $C_o$ = 4.022mm/μs and s = 1.26). Both loading profiles exhibit a slight inflection in the shock front at ~0.1 mm/μs, potentially indicative of the HEL. Although the profile L exhibits some rounding and increases slightly with time, the stress pulse remains within 95% of the peak stress for ~0.5μs, before starting to release. In contrast, profile S exhibits a sharp transition at peak stress and remains within 95% of the peak stress for less than 0.05μs. Upon release both profiles exhibit a classic pull-back signal and ringing, generally indicative of spall or dynamic fracture occurring within the sample[3, 4].

A more detailed view of the pull-back regions are shown in Fig 3.(b). The drops in the free surface velocity (ΔFSV) from the peak states to the minima are 0.056 and 0.067mm/μs for the profiles L and S, respectively. From these results, the spall strengths ($\sigma_{spall}$) for the two experiments can be determined using the relationships proposed by Novikov[42] and the correction proposed by Kanel[43]:

$$\sigma_{spall} = \frac{1}{2}\rho_0 c_B (\Delta FSV + \delta), \quad (1)$$

where $\rho_o$ is the ambient density (19,260kg/m$^3$), $c_B$ is the bulk sound speed (4.022mm/µs), and $\Delta FSV$ is the observed pull-back signal (as shown in Fig 3.(b)). The accuracy of the spall strength is improved by correcting for the $\Delta FSV$ in Equation (1)[43]

$$\delta = h\left(\frac{1}{C_B} - \frac{1}{C_L}\right) * \frac{|\dot{u}_1 * \dot{u}_2|}{|\dot{u}_1| + \dot{u}_2}, \quad (2)$$

where $h$ is the thickness of the spalled region (measured in optical micrographs as ~1.75mm for profile L and 0.4mm for profile S), $C_L$ is the longitudinal sound speed (5.22mm/µs), $\dot{\mathbf{u}}_1$ and $\dot{\mathbf{u}}_2$ are the unloading and re-compression rates calculated as

$$|\dot{u}_1| = -\frac{1}{2}\frac{dFSV}{dt} \quad \text{and} \quad \dot{u}_2 = \frac{1}{2}\frac{dFSV}{dt}, \quad (3)$$

where $\frac{dFSV}{dt}$ is measured from the pull-back signal as shown in Figure 4.(a) and listed in Table II. The unloading rates $\dot{\mathbf{u}}_1$ are normally interpreted as indicators of the kinetics of the tensile pulse imposed on the target [6-7]. As such, these values indicate that the sample subjected to profile S experienced a faster tensile stress rate as compared with the sample subjected to profile L.

**Table II:** Calculated and measured parameters from FSV data

| Exp. ID. | Pull-back characteristics | | | δ (mm/µs) | Spall strength (GPa) | Corrected Spall strength (GPa) |
|---|---|---|---|---|---|---|
| | $\Delta FSV$ (mm/µs) | $\dot{u}_1$ (mm/µs$^2$) | $\dot{u}_2$ (mm/µs$^2$) | | | |
| L | 0.056 | 0.144 | 0.036 | 0.0029 | 2.16 | 2.28 |
| S | 0.067 | 0.295 | 0.114 | 0.0020 | 2.59 | 2.66 |

The corrected spall strength values calculated using Equations 1-3 are listed in Table II, they differ by only ~14 %, being slightly higher in profile S as compared to the profile L. These spall strength values are consistent with those reported in the literature [30, 32, 34]. It is worth noting that the difference between the two specimens is less than the sample-to-sample scatter generally reported for WHA and is consistent with the statistical variation within a given sample reported by Vogler and Clayton[34], therefore being statistically insignificant for this brittle material. Moreover, this difference is an order of magnitude less than reported for the spall strength dependence on wave profile observed in ductile metals[10-12]. For these two reasons the difference between the two samples reported in this article can be considered negligible.

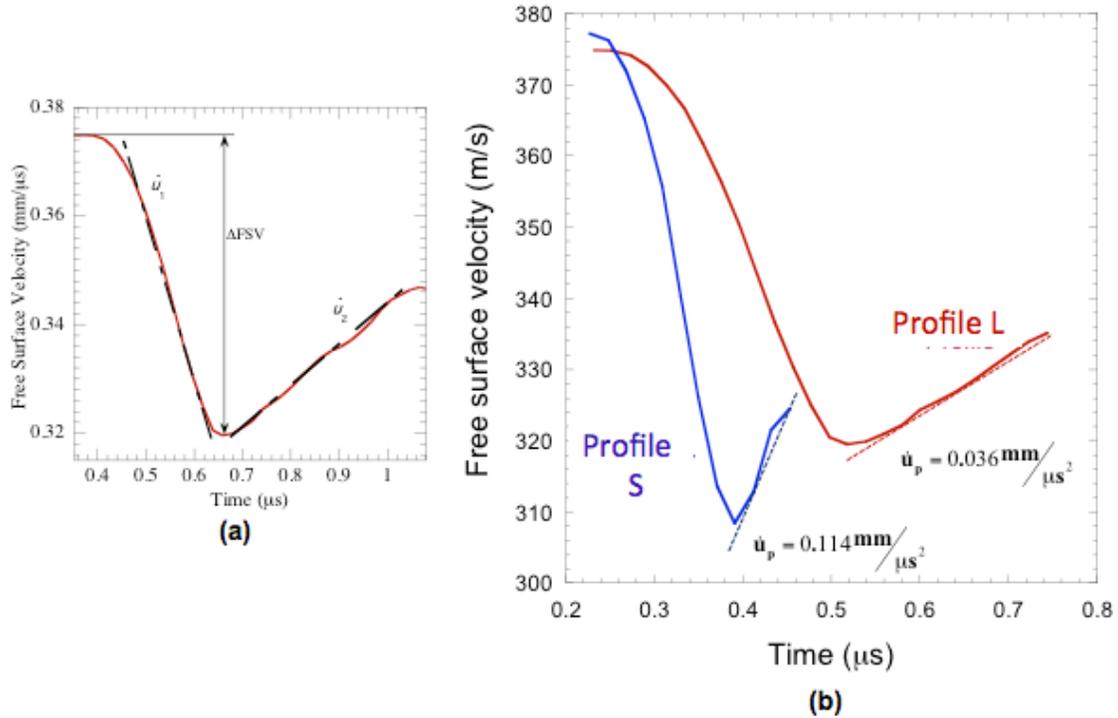

**Fig. 4** (a) Plot of a pull-back signal showing the unloading rate and the re-compression rates used for the correction in Eqs. 1-3. (b) Pull-back signal of the two loading profiles showing the difference in re-compression rates.

Nevertheless, differences are observed in the rate at which the free surface velocity rises beyond the minima, as shown in Fig. 4.(b). The rate for the sample subjected to the profile S is approximately 3 times higher than the rate of the sample subjected to other profile. It has been observed in a variety of materials, both ductile [44-46] and brittle[47], that the rate at which the velocity rises correlates directly to the rate at which the damage develops. As such, higher rates of velocity rise after the velocity minima correlate with a more rapid completion of the damage, or in this case, complete spall fracture. Based on the measurements shown in Fig. 4.(b) a more brittle-like (catastrophic) damage is expected in the sample subjected to profile S as compared to the other loading. A thorough characterization of the spalled samples is presented in the next section to substantiate these assertions.

**B. Post-impact examination**
*i. Optical microscopy*

Several differences are observed in the optical micrographs of the cross sections of the recovered samples presented in Figs. 5.(a)-(d). The location of the spall plane within the thickness of the target is as expected, centered for the profile L but near the rear of the target for profile S. This results from the difference in timing for the interaction of the release waves off the back of the impactor and target because of the different impactor designs, as shown previously in Fig. 2. Of greater interest is that the spall plane in both samples is very localized in the form of cracks consistent with brittle fracture in contrast to the diffuse damage zones seen in ductile metals failing via void nucleation and growth[12]. Minimal plastic deformation is observed in the microstructure adjacent to the spall plane and no voids are seen to occur, as is associated with incipient spall in ductile materials.

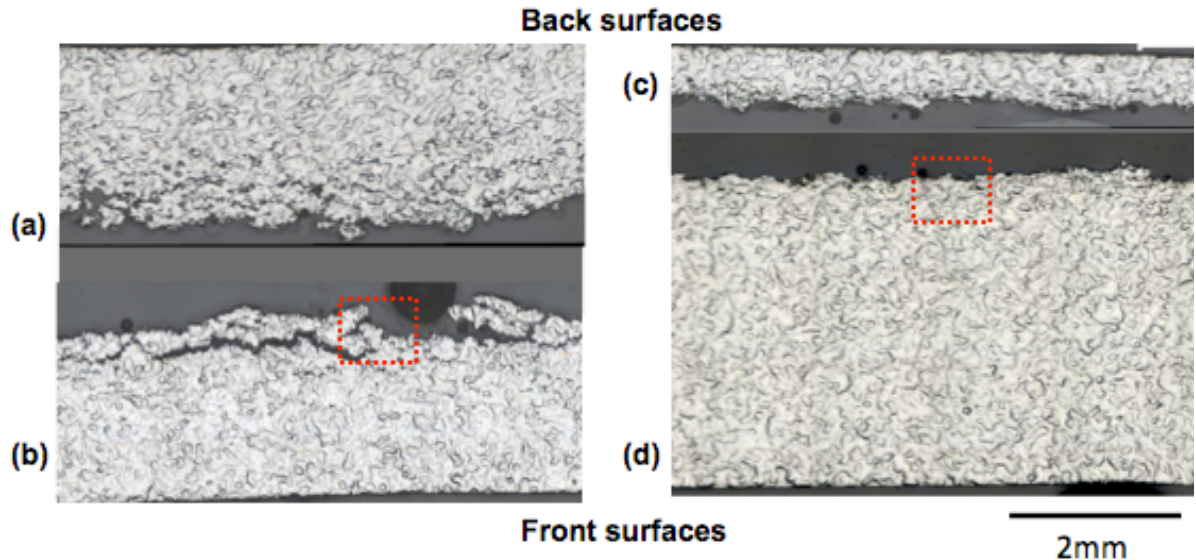

**Fig. 5** Optical micrographs of the spalled samples loaded with (a-b) profile L; and (c-d) profile S. The shock direction is from bottom to top. The spall plane is significantly rougher for the profile L, while it is more localized with less bifurcation of cracks for the profile S.

In neither case are the cracks perfectly flat, but rather follow a path with fracture roughness and crack-path tortuosity consistent with the length-scale of the tungsten particles. A more detailed view of selected areas, signaled by the dotted squares in Figs. 5.(b) and (d), are shown in Figs. 6.(a) and (b). To complement these observations, the results of the profilometry measurements performed on both fracture surfaces are presented in Figs. 6.(c). By loading with a profile L, a wide tensile pulse is imposed within the WHA that causes a rougher or more circuitous spall plane (maximum undulation of the crack path is ~240μm or three to five particle diameters), with the formation of secondary cracks parallel to the primary spall plane. Conversely the damage in the spall plane for the sample subjected to a narrower and faster tensile pulse exhibits a more localized damage with less bifurcation of cracks as they propagate across the spall plane of the sample (maximum undulation height of the crack path is ~90 μm or two particle diameters, with neighboring peaks and troughs typically having significantly lower amplitude).

A lower bound for the size of the tensile pulse can be calculated by the considering the temporal width as the time the free surface velocity is greater than the minimum velocity of the pull-back. Under these considerations, the spatial width results in ~280μm for loading profile L and ~80μm for loading profile S, both values approaching the measured crack path. The fact that crack undulations take up the full spatial width of the shock wave further indicate that any kinetics are completed on a timescale shorter than the time of the shock wave. In this brittle material the time for crack nucleation is negligible compared to the duration of the experiment and the crack propagation rate is limited to the sound speed of the material or in the case of the current experiments by the shock velocity.

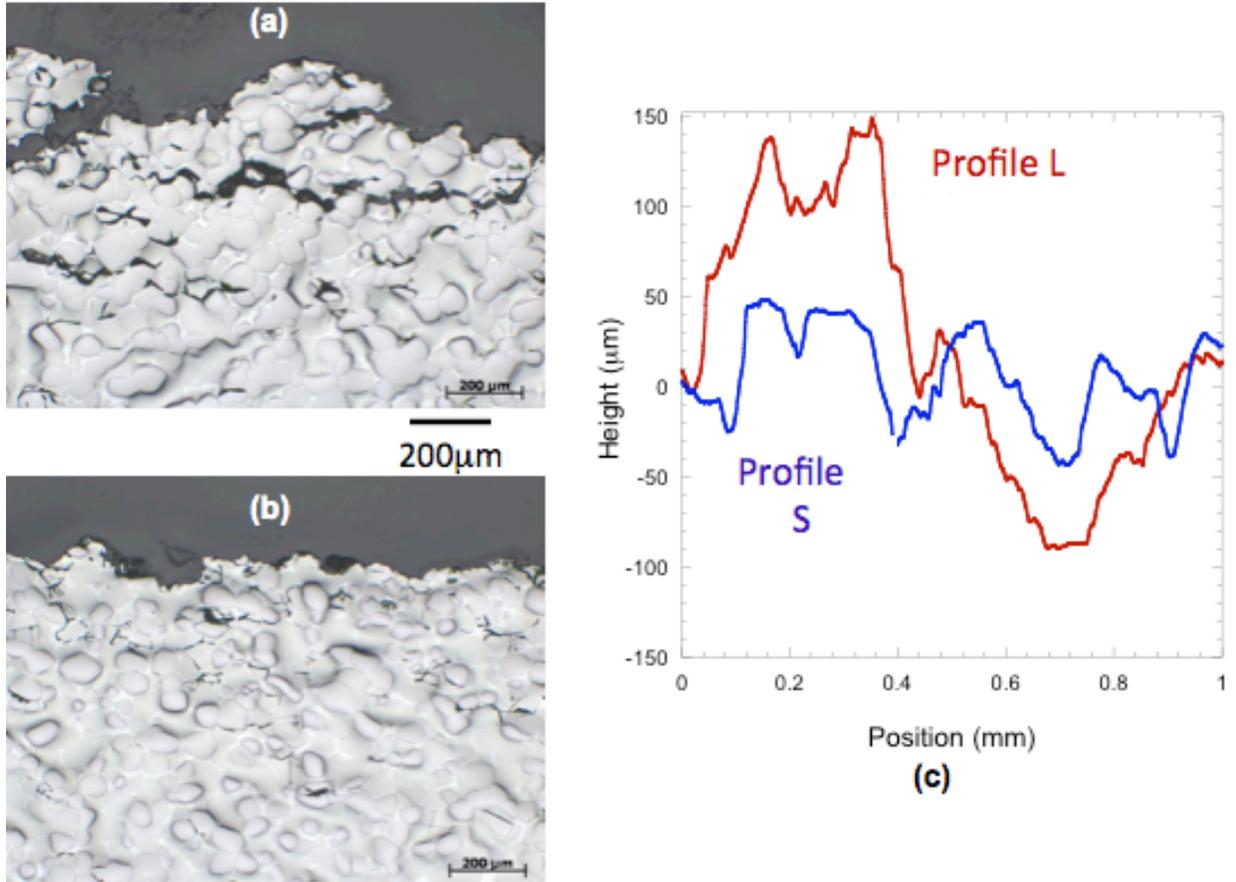

**Fig. 6** Higher magnification optical micrographs of the selected areas in Fig. 5: (a) loading profile L; and (b) loading profile S. The shock direction is from bottom to top. (c) Roughness measured on the fracture surfaces of the recovered samples.

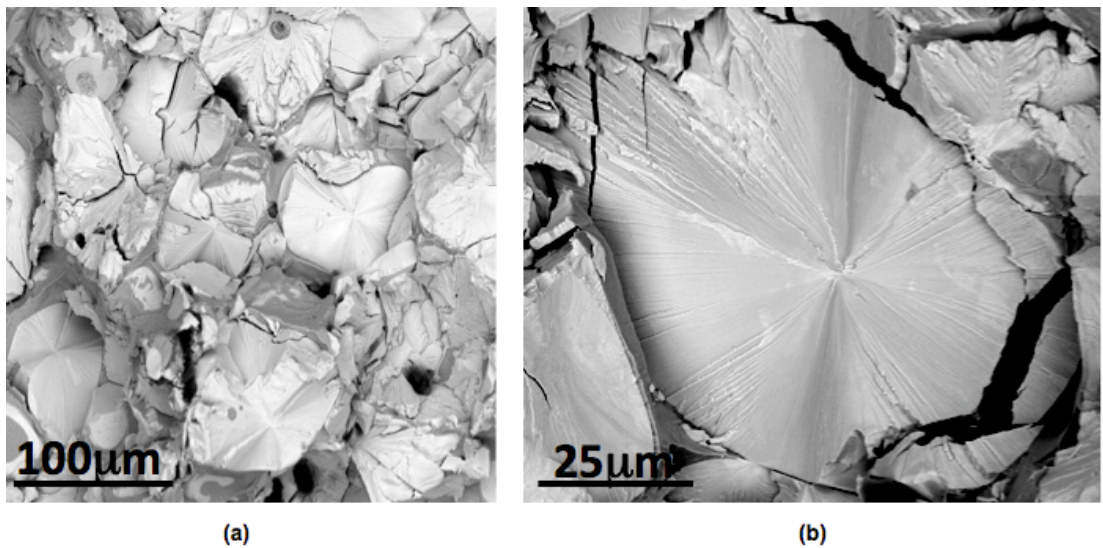

**Fig. 7** SEM images of the fracture surface showing cleavage as the preferred fracture mode for particle cracking: (a) low magnification showing cracked W particles as light and surrounding matrix as dark and (b) higher magnification showing one cracked W particle in detail.

*ii. Scanning electron and electron backscatter diffraction microscopy*

To gain further insight into the fracture modes of this material, a low magnification SEM image of the fracture surface is shown in Fig. 7.(a) and a higher magnification in Fig. 7.(b). Both images show that cracks propagated by cleaving the tungsten particles and linking up through the matrix. In addition to the primary crack plane through the particles, many of the particles exhibit extensive secondary cracking (Fig. 7.(b)), a mechanism for further dissipating energy.

Electron backscatter diffraction (EBSD) images in Figs. 8.(a) and (c), further highlight secondary cracks formed underneath the primary crack (i.e. spall plane). As observed in Fig. 8.(a), there is a set of cracks that have linked up across multiple particles for the loading profile L (indicated by the white arrow). Additionally, both samples exhibit significant fracturing of individual W particles where the crack does not appear to propagate into the matrix or surrounding particles. This type of sub-surface damage of the brittle phase is often observed when mechanical toughening is achieved through a multi-phase composite microstructure[50].

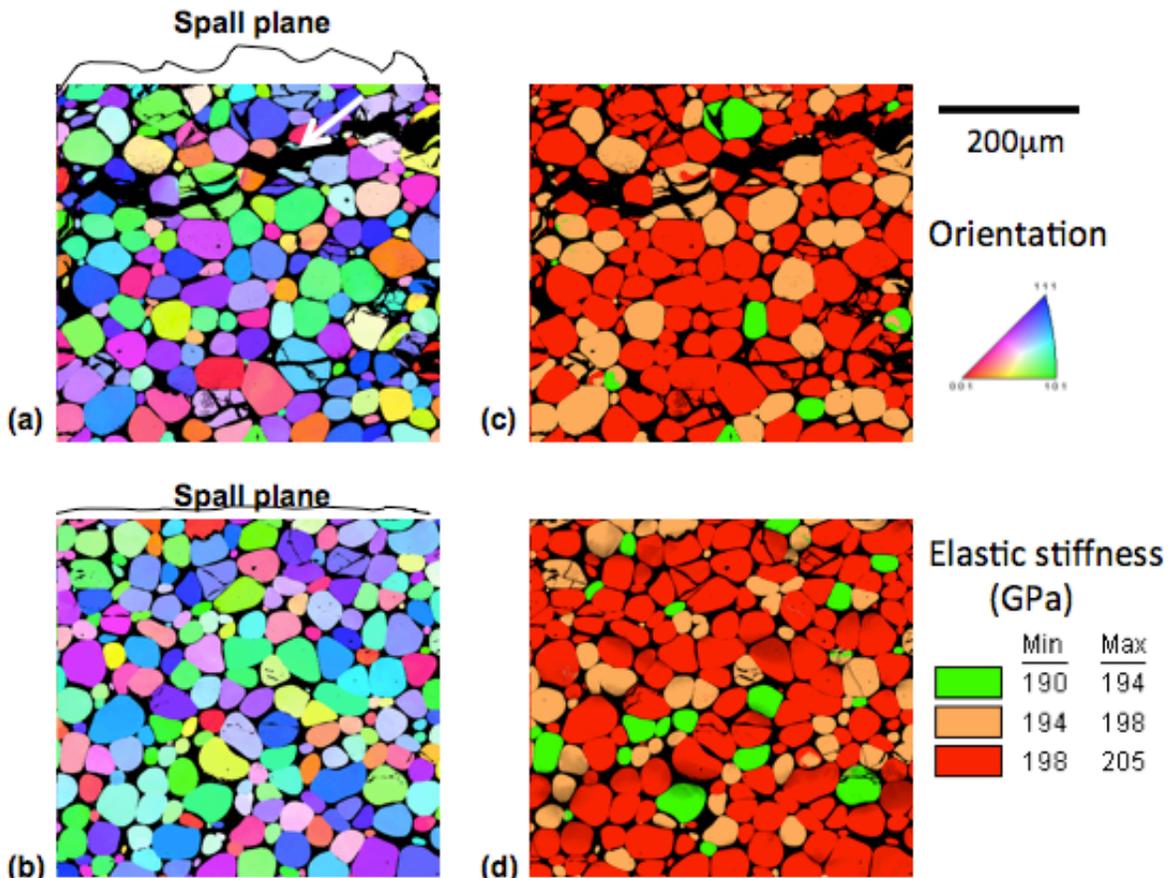

**Fig. 8** Orientation maps near the spall plane of the spalled specimens loaded with: (a) profile L and (b) profile S. The shock direction is from bottom to top and the color indicates the particle's crystalline orientation with respect to the shock direction according to the color key. (c-d) Elastic stiffness maps of the same areas. For each particle the elastic stiffness is calculated based on the particle's crystallographic orientation and the loading conditions. In these maps the color is assigned according to the ranges given in the list.

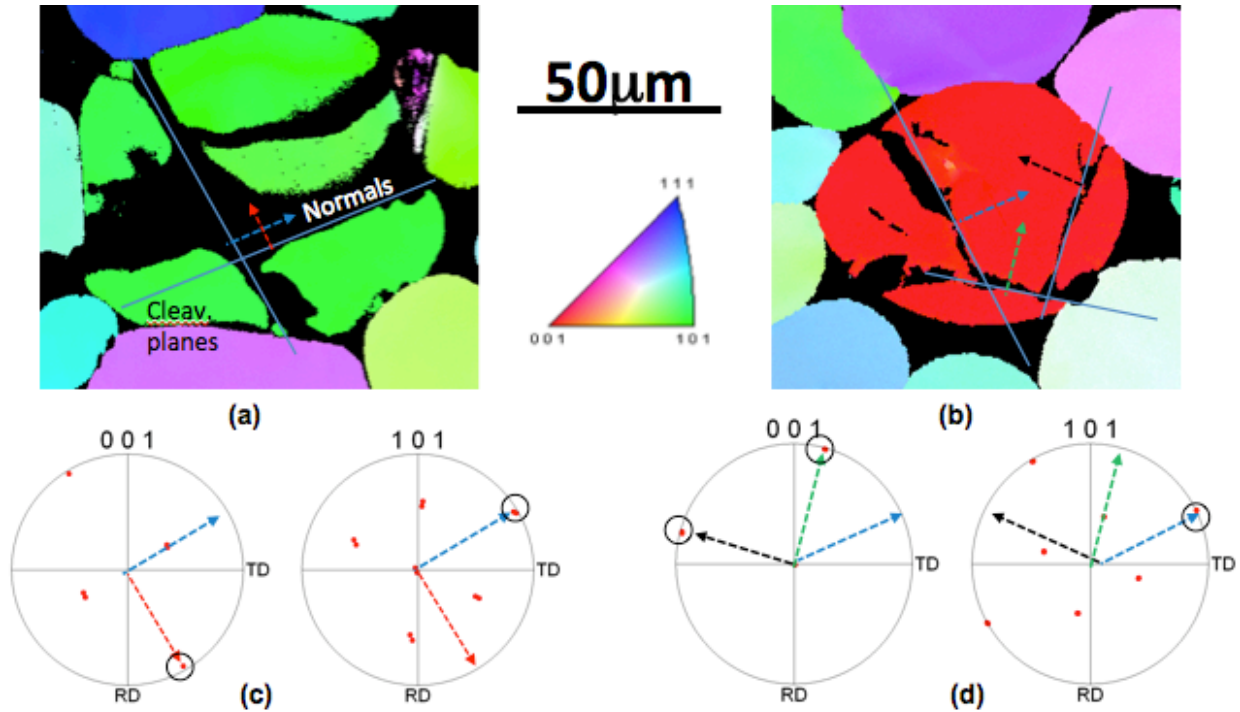

**Fig. 9** Orientation maps showing two cracked particles of the samples loaded with (a) Profile L and (b) Profile S. The superimposed arrows indicate the normal of the planes of the cracks, or cleavage planes. (c-d) Pole figures showing the projections of the same normals as in (a) and (b). In both spalled samples the cleavage planes coincide with {0 0 1} and {1 0 1} planes.

Elastic stiffness maps generated from EBSD data are shown in Figs. 8.(c) and (d). This property is calculated by considering the crystallographic orientation and loading conditions. As shown in Figs. 8.(c) and (d), cracking preferentially occurred in the particles with the highest elastic stiffness in the direction of the shock propagation (colored as red). These results can be explained in terms of the uniaxial strain conditions present in shock experiments. If an average strain is imposed on the entire sample, it follows that the particles with higher elastic modulus would develop higher stress concentrations, and as such failure would be promoted in these particles, as observed in these images.

Further analysis using EBSD data is shown in Figs. 9.(a) and (b). The images show detailed views of two given particles containing cracks. By using projections of the normal of a given plane of a crack onto the respective pole figure of the particle (Figs. 9.(c) and (d)), it is observed that the planes of the cracks correspond to the {0 0 1} and {1 0 1} planes. These observations are consistent with the observed preferred orientations for grain cleavage in W[51].

**IV Discussion**

Although the imposed wave profiles do not significantly affect the value of the spall strength of WHA, the difference in wave-form does significantly change the area of the sample put into tension. A wide damage zone in the form of cracks is observed in the sample loaded with profile L as compared to the loaded with profile S. It is noteworthy that this difference in the amount of additional cracking is captured as the acceleration measured in the free surface velocity trace after the pull-back minima (Fig. 4.(b)). The acceleration rates appear to be indicative of the ability to form a complete spall plane. In this regard, the higher rate measured in profile S correlates with a single, flatter, less tortuous, spall plane observed in the spalled sample as compared to the sample subjected to the profile L (Figs. 5 and 6). In both cases, the height of the region of damage corresponds very closely to the spatial width of the release shock waves.

Additionally, the current data indicates that cleavage along W particles with high elastic modulus is the dominant damage evolution mechanism in WHA, although it did not occur along the planes expected by the Gumbsch *et al* [51] study. These observations are, however, consistent with the work by Vogler and Clayton[34]. Through simulations, they hypothesized that crack propagation and macroscopic spall behavior would also require substantial grain cleavage. Once sufficiently large, a micro-crack initiated along the boundary between a particle and matrix would propagate fully across the specimen, irrespective of the underlying microstructure. In agreement with this hypothesis, the fracture surfaces shown in Fig. 7 indicate that both inter-granular and cleavage mechanisms are operative.

Interestingly, isolated fractured W particles located away from the spall region, i.e. the region that sustained the maximum state of tension, are observed. Thus, it is possible that some of these particles may have been fractured during the initial compressive shock-wave loading perhaps at particle-particle intersections where matrix deformation cannot relieve the local imposed shear stresses. A likely scenario is that upon shock loading, the imposed compressive stress generates localized tensile and deviatoric (shear) stresses due to the difference in mass impedance between particle and matrix. The action of these stresses might have additionally caused the particles to crack at locations where the threshold stress for cleavage might have been overcome.

**V Summary**
Plate impact experiments were conducted to examine the effect of loading profile on the dynamic tensile response of tungsten heavy alloy (WHA) specimens. Characterization of the as- received and spalled WHA specimens was performed using optical, SEM and EBSD microscopy. The main findings are as follows: to the shock pulse duration
   and the shock pulse release or unloading,
- Contrary to ductile metals where the kinetics for void initiation, growth, and linkage result in a strong dependence of spall strength on wave profile shape, the dynamic damage of tungsten heavy alloy (WHA) is seen to be relatively insensitive to the to the shock pulse duration and accompanying unloading (tensile) pulse. The difference in spall strength for WHA at ~15.4 GPa was measured to be statistically insignificant (~14%) for the two wave profiles explored in the current work.
- In all cases failure is by brittle cleavage through the tungsten particles and link-up through the matrix resulting in a circuitous path on the length-scale of the particles. The difference in the characteristics of the loading profiles led to differences in the roughness or circuitousness of the crack plane and the initiation of parallel secondary cracks and crack bifurcation. While this enabled the brittle material to dissipate more energy, it did not change the strength of the material.
- Cracking is observed to preferentially occur in particles with high elastic stiffness and along {0 0 1} and {1 0 1} planes.

In conclusion, in this brittle material all relevant damage kinetics and the spall strength are shown to be dominated by the shock peak stress, independent of pulse duration.

**VI Acknowledgements**
Los Alamos National Laboratory is operated by LANS, LLC, for the NNSA of the US Department of Energy under contract DE-AC52-06NA25396. This research was supported under the auspices of the US Department of Energy and the Joint DoD/DOE Munitions Program. The authors would like to acknowledge Mike Lopez and Rob Dickerson for their help in sample prep and some of the data analysis.